\documentstyle[preprint,aps,12pt]{revtex} 
\begin{document} 
\draft 
\noindent 
\begin{centering} 
 
{\Large DYNAMICAL MODEL AND PATH INTEGRAL FORMALISM FOR HUBBARD OPERATORS.\\} 
 
\vspace{1.cm} 
{\bf A. Foussats, A. Greco and O. S. Zandron}\\ 
 
\vspace{1cm} 
 
{\em Facultad de Ciencias Exactas Ingenier\'{\i}a y Agrimensura e IFIR  
(UNR-CONICET)\\ 
Av.Pellegrini 250 - 2000 Rosario - Argentina.}\\ 
 
\end{centering} 
 
\begin{abstract} 
 
\begin{centering} 
{\bf Abstract\\} 
\end{centering} 
 
In this paper, the possibility to construct a path integral formalism  
by using the Hubbard operators as field dynamical variables is investigated. 
By means of arguments coming from the Faddeev-Jackiw symplectic Lagrangian 
formalism as well as from the Hamiltonian Dirac method, it can be shown  
that it is not possible to define a classical dynamics  
consistent with the full algebra of the Hubbard $X$-operators. 
Moreover, from the Faddeev-Jackiw symplectic algorithm, and in order to  
satisfy the Hubbard $X$-operators commutation rules, it is 
possible to determine the number of constraint that must 
be included in a classical dynamical model. Following this approach it remains  
clear how the constraint conditions that must be introduced in the classical  
Lagrangian formulation, are weaker than the constraint conditions imposed  
by the full Hubbard operators algebra. The consequence of this fact is  
analyzed in the context of the path integral formalism. Finally, in the  
framework of the perturbative theory, the diagrammatic and the Feynman rules  
of the model are discussed.  
	    
\end{abstract} 
 
 
\narrowtext 
  
\section{Introduction}

The Hubbard $X$-operators\cite{1} are suitable to give a  
powerfull framework in which the elementary excitations in solids can be  
explained.  
The use of $X$-operators is also relevant when electronic correlations are 
taken into account. This is the scenery in which High-$T_{c}$ 
superconductivity occurs, and so the main reason why the 
Hubbard operators algebra is so interesting at the present time. 
   
The algebra of the Hubbard ${\hat{X}}$-operators is completely defined by:  
 
a) the commutation rules 
 
\begin{equation} 
[{\hat{X}}_{i}^{\alpha \beta}\,,\,{\hat{X}}_{j}^{\gamma \delta}] 
=\delta_{ij}(\delta^{\beta \gamma}{\hat{X}}_{i}^{\alpha \delta} - 
\delta^{\alpha \delta}{\hat{X}}_{i}^{\gamma \beta})  
\eqnum{1.1} 
\end{equation}

b) the completness condition 
 
\begin{equation} 
{\hat{X}}_{i}^{+ +} + {\hat{X}}_{i}^{- -} = {\hat {I}} 
\eqnum{1.2} 
\end{equation} 
 
c) the multiplication rules for a given site 
 
\begin{equation} 
{\hat{X}}_{i}^{\alpha \beta} {\hat{X}}_{i}^{\gamma \delta} = 
\delta^{\beta \gamma} {\hat{X}}_{i}^{\alpha \delta}\;. 
\eqnum{1.3} 
\end{equation} 
    
From now on and for simplicity  we consider the case in which  
the indices $\alpha,\beta$ can only take the values $+$ and $-$, and so  
the 
Hubbard ${\hat{X}}$-operators are boson-like operators of the $SU(2)$ algebra.  
The spin $s = 1/2$ is naturally contained in this case. 
 
It is easy to show that the equations (1.3) are not all independent, 
and so the full information contained in the algebra can be recovered from 
the equations (1.1), (1.2) and the following three independent equations       
 
\begin{equation} 
{\hat X}^{- +}{\hat X}^{+ +} - {\hat X}^{- +} = 0 
\eqnum{1.4a} 
\end{equation} 
 
\begin{equation} 
{\hat X}^{+ -} {\hat X}^{- -} - {\hat X}^{+ -} = 0 
\eqnum{1.4b} 
\end{equation} 
 
\begin{equation} 
{\hat X}^{+ -} {\hat X}^{- +} - {\hat X}^{+ +} = 0   
\eqnum{1.4c} 
\end{equation} 
 
Consequently, the full algebra given by equations (1.1)-(1.3) is equivalent  
to the commutation rules (1.1), the completness condition (1.2) and the three 
conditions (1.4).  
 
A many body theory constructed by using the Hubbard operators as field 
variables, requires the application of techniques used in quantum field  
theories. From this point of view it is necessary to formulate  
the Wick theorem for the case in which the field operators are neither usual  
fermion nor bosons. Progress in this directions where done\cite{2}, but the 
problem is still open. 
 
Like in quantum field theories, another way to attach the problem is via 
the path integral formulation. It is important to say that a suitable path 
integral formulation must be independent of a given representation. On the 
other hand it must be written in terms of an effective action with a well 
defined dynamics. This last point of view will be adopted in the present  
paper.    
 
The paper is organized as follows. In section II and III, by using the  
Faddeev-Jackiw (FJ) Lagrangian method\cite{3}, a general  
treatment for first-order Lagrangian systems containing the Hubbard  
operators as dynamical variables is given. A family of Lagrangian describing 
these dynamical systems is found. The use of these classical Lagrangians 
in a path integral quantization formalism is also analyzed. Strong arguments  
can be given showing that it is not possible to include the full  
Hubbard algebra (1.1)-(1.3) in a classical dynamical model. 
In section IV, we confront our results with others previously given in the  
literature. In section V, the diagrammatic and the Feynman rules for the  
model are constructed. Finally, conclussions and discussions are given in  
section VI. 
 
\section{Classical Lagrangian and Dynamical Model}    
 
One of the traditional approaches to study the quantization of spin systems 
or t-J model in which the Hubbard operator algebra takes place, is to  
consider the constrained systems from the point of view of coherent state 
phase path integration. Also the usual Dirac's Hamiltonian method for 
constrained systems by considering slave boson or fermion representation 
is frequently used.  
  
By writing a family of first-order classical Lagrangian directly in terms 
of the four Hubbard operators, our main purpose 
is to obtain information about the kind and the number of constraints present 
in these models. In this way it is possible to obtain a response 
about how many information contained in the algebra (1.1)-(1.3) can be 
introduced at the classical level. This approach requires the introduction  of 
a suitable set of constraints, a priori unknown, that must be determined 
later on.   
To this purpose it is useful to use the FJ Lagrangian method\cite{3,4,5,6}.  
Therefore, we briefly indroduce some definitions and key equations.  
 
As is well known, the FJ symplectic quantization method is formulated on 
actions only containing first-order time-derivatives.  
The most general first-order Lagrangian is specified in terms of two  
arbitrary functionals $K_{A}(\mu^{A})$ and ${\bf V^{(0)}}(\mu)$,  
and is given by 
 
\begin{equation} 
L(\mu_{A}\,,\,{\dot{\mu}^{A}}) = {\dot{\mu}}^{A}K_{A} 
(\mu^{A}) - {\bf V^{(0)}}(\mu) \; .  
\eqnum{2.1} 
\end{equation} 
 
The functionals, $K_{A}(\mu^{A})$ are the components of the canonical  
one-form $K(\mu) = K_{A}(\mu)d{\mu^{A}}$ and the functional  
${\bf V^{(0)}}(\mu)$ is the  
symplectic potential. The general compound index $A$ runs in the 
different ranges of the complete set of variables that defines the  
extended configuration space. 
 
The Euler-Lagrange equations of motion obtained from (2.1) are: 
 
\begin{equation} 
\sum_{B}M_{AB}\dot{\mu}^{B} - \frac{\partial {\bf V^{(0)}}} 
{\partial \mu^{A}} \;=\;0\;. 
\eqnum{2.2} 
\end{equation}  
 
The elements of the symplectic matrix $M_{AB}(\mu)$ are the components of 
the sympletic two-form $M(\mu) = dK(\mu)$. The exterior derivative of 
the canonical one-form $K(\mu)$ is written as the generalized curl 
constructed with partial derivatives and so, the components are given by: 
    
\begin{equation} 
M_{AB} = \frac{\partial K_{B}}{\partial \mu^{A}} -  
\frac{\partial K_{A}}{\partial \mu^{B}}  
\;.  
\eqnum{2.3} 
\end{equation} 
 
When the symplectic matrix $M_{AB}$ is non-singular,  
from the equations of motion (2.2) result 
 
\begin{equation}  
\dot{\mu}^{A} = (M^{AB})^{-1} \frac{\partial{\bf V^{(0)}}} 
{\partial{\mu}^{B}} 
\eqnum{2.4} \,. 
\end{equation} 
 
As the symplectic potential is just the Hamiltonian of the system,  
the equation (2.4) is written 
 
\begin{equation}  
\dot{\mu}^{A} =  \left[{\mu}^{A}\,,\,{\bf V}\right] =  
\left[{\mu}^{A}\,,\,{\mu}^{B}\right]\frac{\partial {\bf V^{(0)}}} 
{\partial{\mu}^{B}} 
\eqnum{2.5} \,, 
\end{equation} 
 
\noindent 
where 
  
\begin{equation}  
\left[\mu^{A}\,,\,\mu^{B}\right] =  
(M^{AB})^{-1} 
\eqnum{2.6}\, , 
\end{equation} 
 
\noindent 
are the generalized brackets defined in the FJ symplectic formalism. 
 
It is easy to show that the elements $(M^{AB})^{-1}$ of the inverse   
of the symplectic matrix $M_{AB}$ 
correspond to the Dirac brackets\cite{7} of the theory. 
   
Transition to the quantum theory is realized as usual replacing classical  
fields by quantum field operators acting on the Hilbert space, 
where quantum ordering and proper  
boundary conditions for the quantum field operators must be taken into account. 
Therefore, the predictions of both FJ and Dirac methods are equivalents. 
  
When the matrix $M^{AB}$ is singular,  
the  constraints appear as algebraic relations and they 
are necessary to maintain the consistency of the field equations of motion.    
In such a case, there exist $m$ ($m<n$) left (or right) zero-modes ${\bf  
v}_{a}$ 
($a = 1,...,m\,,\,A = 1,...,n$) of the supermatrix $M_{AB}$, where each  
${\bf v}_{a}$ is a 
column vector with $n+m$ entries $v^{A}_{a}$.  
So the zero-modes verify the following equation  
 
\begin{equation} 
\sum_{A} v^{A}_{a} M_{AB} = 0 \; .  
\eqnum{2.7} 
\end{equation} 
 
From the equations of motion (2.2) we see that the quantities  
$\Omega_{a}$ are the true  
constraints in the FJ symplectic formalism, and they are given by 
  
\begin{equation}  
{\Omega_{a}} =  v^{i}_{a}\;\frac{\partial} 
{\partial \varphi^{i}} {\bf V^{(0)}} = 0 \;. 
\eqnum{2.8} 
\end{equation}

Consequently, in a first iteration the constraints are written in the 
symplectic part of the Lagrangian by means of Lagrange multipliers as follows   
 
\begin{equation} 
L^{(1)} = {\dot{\varphi}}^{i}a_{i}(\varphi)  
+ {\dot{\xi}}^{a} \Omega_{a} - {\bf V}^{(1)}  
\hspace{1cm} \; ,  
\eqnum{2.9} 
\end{equation} 
 
\noindent 
where the new symplectic potential is by definition  
${\bf V}^{(1)} = {\bf V}^{(0)}{\mid}_{{\Omega} = 0}$.  
The partition $\mu^{A} = (\varphi^{i}\,,\,\xi^{a})$ and $K_{A} = (a_{i}\,,\, 
\Omega_{a})$ has been made. So, the compound indices $A,B$ run the set 
$A = (i\,,\,a)$ and $B = (j\,,\,b)$. 
 
In each iterative procedure the configuration space is enlarged and the 
symplectic matrix is modified. When no new constraints are found the 
iterative procedure is finished.  
 
Now, we are going to apply the FJ quantization formalism to a dynamical 
model for the Hubbard operators. 
 
As it is well known in all the examples in which  
the field variables are the components of the spin operators, the starting  
point is to consider first-order Lagrangians. 
This also happens in the t-J model when it is written in slave boson or  
fermion  
representation\cite{8}. The FJ quantization algorithm is suitable to 
study this kind of dynamical systems described by constrained  
first-order Lagrangians in which the constraints play a crucial 
role.  
 
Therefore, in the case under consideration we assume that the first-order  
classical Lagrangian as functional of the Hubbard operators is written as  
follows 
 
\begin{equation} 
L = a_{\alpha \beta}(X){\dot{X}}^{\alpha \beta} - H(X) - \lambda_{a} 
\Omega^{a} \;, 
\eqnum{2.10} 
\end{equation}

\noindent 
where $H(X)$ is for instance the Hamiltonian of the Heisenberg model written  
in terms of the Hubbard operators. The site indices were dropped since 
they are irrelevant in the analysis we will develop. Without any 
difficulty the site indices can be opportunely included.  
 
In the equation (2.10) $\lambda_{a}$ is an adequate set of Lagrange  
multipliers which allows the introduction of the constraints in the  
Lagrangian formalism. $\Omega^{a}(X)$ is a set of suitable unknown  
constraints,  
initially considered ad hoc in the Lagrangian. Both the constraints  
$\Omega^{a}(X)$ as well as the range of the index $a$	 
must be determined by consistency.  
The coefficients  
$a_{\alpha \beta}(X)$ = $a^{*}_{\beta \alpha}(X)$ are found in      
such a way that the algebra (1.1)-(1.3) for the Hubbard operators must be   
verified. 
 
Looking at equation (2.10) we see that   
the initial set of dynamical symplectic variables 
is defined by $(X^{\alpha \beta}\,,\,\lambda_{a})$ and  
the symplectic potential ${\bf V^{(0)}}$ is given by 
 
\begin{equation} 
{\bf V^{(0)}} = H(X) + \lambda_{a}\Omega^{a}\;.  
\eqnum{2.11} 
\end{equation} 
 
So, the symplectic matrix (2.3) obtained from the Lagrangian (2.10) is  
singular,  
therefore the constraints are obtained by using the equation (2.8) and they 
read

\begin{equation} 
\frac{\partial{\bf V}^{(0)}}{\partial{\lambda_{a}}} = \Omega^{a}\;, 
\eqnum{2.12} 
\end{equation}

\noindent 
and the first-iterated Lagrangian writes 
 
\begin{equation} 
L^{(1)} = a_{\alpha \beta}(X){\dot{X}}^{\alpha \beta} + {\dot{\xi}_{a}} 
\Omega^{a} - H(X)\;.   
\eqnum{2.13} 
\end{equation} 
 
The modified symplectic matrix associated to the Lagrangian (2.13) is 
 
\begin{eqnarray} 
M_{AB} = 
\left( \matrix { 
\frac{\partial a_{\gamma \delta}}{\partial X^{\alpha \beta}} 
- \frac{\partial a_{\alpha \beta}}{\partial X^{\gamma \delta}} 
&\frac{\partial\Omega_{b}}{\partial X^{\alpha \beta}}   \cr 
- \frac{\partial\Omega_{a}}{\partial X^{\gamma \delta}}    &0    \cr } 
\right) \; , \eqnum{2.14} 
\end{eqnarray} 
 
\noindent 
where the indices $A = \{{(\alpha \beta)}, a \}$ ,$B =\{{(\gamma \delta)}, 
b\}$. 
  
At this stage the problem is to determine which, and how many constraints  
can be deduced from the algorithm of the method in such a way  
to obtain a non-singular symplectic matrix. 
 
In this way, from the Lagrangian (2.13) the symplectic matric (2.14) is 
constructed and its inverse can be computed. By equating each elements  
$(M^{AB})^{-1}$ of the inverse of the symplectic matrix $M_{AB}$  
to each one of the commutations rules (1.1), differential equations on the   
coefficients $a_{\alpha \beta}(X)$ and on the constraints $\Omega^{a}$ 
are obtained. 
  
As it can be seen, the dimension of the symplectic matrix (2.14) is 4 + $a$, 
where $a$ enumerates the constraints. Because of the antisymmetric 
property of $M_{AB}$ the index $a$ has even range.  
From the properties of this matrix 
we can conclude:  
 
\noindent 
i) If $a>4$ or odd, the symplectic matrix is singular.  
 
\noindent 
ii) For $a = 4$ the symplectic matrix can be invertible, but it is not  
possible to obtain the commutation rules (1.1). The commutators obtained by  
using equation (2.6) vanish, independently of the value of the coefficients  
$a_{\alpha \beta}(X)$. On the other hand, when the 
number of constraints equals the number of fields there is no dynamics.    
So, it is not possible by means of Lagrange multipliers to enforce  
the constraint (1.2) together with the other three conditions (1.4). 
 
Consequently, we must resign the introduction in a classical first-order  
Lagrangian of the complete information contained in the algebra (1.1)-(1.3).  
 
Then, the unique possibility is to have only two constraints. 
The equation (1.2) or completness condition must be imposed  
accounting their physical meaning.  
It avoids at quantum level the configuration with doubly occupied sites.   
The remaining constraint can not be 
any one of that given in (1.4), because the commutators (1.1) can not be 
recovered. 
Therefore, we can expect that the remaining constraint  
can be provided naturally by consistency, when the symplectic method is used. 
   
Consequently, we assume an arbitrary constraint  
$\Omega = \Omega\,(\,X^{+ -}\,,\, 
X^{- +}\,,\,u\,)$, where $u\, = \,X^{+ +} - X^{- -}$.  
This assumption is not a restriction because, by the completness condition, 
the sum \, $(X^{+ +} + X^{- -})$ is equal to one.   
From the requirement that the matrix elements of the  
inverse of the symplectic matrix (2.14) must be equal to each one of   
the Hubbard commutation rules (1.1), and by solving the differential 
equation on this constraint the solution we find is  
 
\begin{equation} 
\Omega_{} = X^{+ -} X^{- +} + \frac{1}{4} u^{2} - \beta = 0\; , 
\eqnum{2.15} 
\end{equation} 
 
\noindent 
where $\beta$ is an arbitrary constant. 
 
We emphasize that the constraint (2.15) is not an imposition but appears 
naturally from our method. This is the unique possible constraint in order to 
satisfy the commutation rules and the completness condition. Of course 
in equation (2.15) there is less dynamical information than  
the contained in the three equations (1.4).  
 
We will discuss 
this point connected to the fact that the path integral for this kind of 
fields represents the system in some limit of the operatorial approach.

\section{Determination of the Lagrangian coefficients.}  
 
The next step is to determine the functions $a_{\alpha \beta}(X)$  
written in the Lagrangian (2.13). The two constraints $\Omega_{a}$ 
we must consider are given in equations (1.2) and (2.15). Once the symplectic  
matrix 
(2.14) is constructed its inverse can be computed. Taking into account  
the equation (2.6) and the commutation rules (1.1), by consistency  
the following differential equation is found, 
 
\begin{equation} 
2 [\frac{\partial a_{+ -}}{\partial u} - \frac{\partial a_{u}}{\partial 
X^{+ -}}]\; X^{+ -}  
- 2 [\frac{\partial a_{- +}}{\partial u} - \frac{\partial a_{u}}{\partial 
X^{- +}}]\; X^{- +}  
+ [\frac{\partial a_{- +}}{\partial X^{+ -}} - \frac{\partial a_{+ -}}{\partial 
X^{- +}}]\; u = i \;, 
\eqnum{3.1} 
\end{equation} 
 
\noindent 
where $a_{u} = \frac{1}{2}(a^{+ +} - a^{- -})$. 
 
We assume that the coefficients $a_{+ -}\,,\,a_{- +}$ and $a_{u}$ can be  
written as products of arbitrary functions of the $u$ variable by  
polynomials in the $X^{+ -}$ and $X^{- +}$ variables . 
For simplicity we try to look for a particular family of solutions  
by taking first-order polynomials in the  $X^{+ -}$ and  
$X^{- +}$ variables, i.e    
 
\begin{equation} 
a_{+ -} = f(u) [e + b X^{+ -} + c X^{- +}]\; , 
\eqnum{3.2a} 
\end{equation} 
 
\begin{equation} 
a_{- +} =  a^{*}_{+ -} = f^{*}(u) [e^{*} + c^{*} X^{+ -} + b^{*} X^{- +}]\; , 
\eqnum{3.2b} 
\end{equation}

\begin{equation} 
a_{u} = h(u) [p + q X^{+ -} + r X^{- +}]\; , 
\eqnum{3.2c} 
\end{equation} 
 
\noindent 
where the constant coefficients $p\,,\,q\,,\,r\,,\,e\,,\,b$ and $c$  
are arbitrary ones. 
 
Once the expressions (3.2) are introduced in the equation (3.1) by  
straightforward computation we find 
  
\begin{equation} 
p h(u) = (p h(u))^{*}  
\eqnum{3.3a} 
\end{equation} 
 
\begin{equation} 
q h(u) = (r h(u))^{*}  
\eqnum{3.3b} 
\end{equation}

\begin{equation} 
q h(u) = e \frac{df}{du}  
\eqnum{3.3c} 
\end{equation} 
 
\begin{equation} 
c f(u) - c^{*} f^{*}(u) = 2i{\bf Im}\;cf = 2i \frac{u + \alpha} 
{4 \beta - u^{2}}   
\eqnum{3.3d} 
\end{equation} 
 
\noindent 
with the conditions $b = 0$, and being $\alpha$ an arbitrary  
integration constant.   
 
Consequently, the equations (3.2) for the Lagrangian coefficients and (3.3),  
determine a family of  
Lagrangians compatible with the commutation rules (1.1), the completness 
condition (1.2) and the constraint (2.15). 
 
Not losing generality, in equations 
(3.2b) and (3.3d) we can choose $c = i$ and the function $f(u)$ results   
 
\begin{eqnarray*} 
f(u) = \frac{u + \alpha}{4 \beta - u^{2}}\;,  
\end{eqnarray*} 
 
\noindent  
and so two different families of solutions are obtained: 
 
\noindent 
i) If $e = 0$ the solution reads    
 
\begin{equation} 
a_{+ -} = i \frac{(u + \alpha)}{(4 \beta - u^{2})} X^{- +}\;, 
\eqnum{3.4a} 
\end{equation} 
 
\begin{equation} 
a_{- +} = - i \frac{(u + \alpha)}{(4 \beta - u^{2})} X^{+ -}\;, 
\eqnum{3.4b} 
\end{equation} 
  
\begin{equation} 
a_{u} = \frac{1}{2}(a_{+ +} - a_{- -}) = h(u) \;, 
\eqnum{3.4c} 
\end{equation} 
 
\noindent 
where $h(u)$ is an arbitrary real function which also can be taken equal 
to zero.   
 
\noindent 
ii) If $e\neq0$ the solution reads

\begin{equation} 
a_{+ -} = \frac{(u + \alpha)}{(4 \beta - u^{2})}(1 +i X^{- +})\;, 
\eqnum{3.5a} 
\end{equation} 
 
\begin{equation} 
a_{- +} = \frac{(u + \alpha)}{(4 \beta - u^{2})}(1 - i X^{+ -})\;, 
\eqnum{3.5b} 
\end{equation} 
  
\begin{equation} 
a_{u} = \frac{1}{2}(a_{+ +} - a_{- -}) = h(u)[1 + X^{+ -} + X^{- +}] \;, 
\eqnum{3.5c} 
\end{equation} 
 
\noindent 
where in this second case $h(u)$ verifies the equation (3.3c). 
 
Really, the two different families of solutions (3.4) and (3.5) take into  
account the majority of the significant cases.  
 
Finally, we note that making the following linear transformation to real 
variables $(S_{1}\,,\,S_{2}\,,\,S_{3})$ 
 
\begin{equation} 
X^{+ -} = S_{1} + iS_{2}\;, 
\eqnum{3.6a} 
\end{equation}

\begin{equation} 
X^{- +} = S_{1} - iS_{2}\;, 
\eqnum{3.6b} 
\end{equation}

\begin{equation} 
X^{+ +} - X^{- -} = 2S_{3} \;, 
\eqnum{3.6c} 
\end{equation} 
 
\noindent 
and by defining the vectors\, $\,{\bf a}\, = \,(\,a_{S_{1}}\,,\,a_{S_{2}}\,,\, 
a_{S_{3}}\,)$\,,\, 
$\,{\bf \nabla}\, = \,(\,\partial_{S_{1}}\,,\,\partial_{S_{2}}\,,\, 
\partial_{S_{3}}\,)$\, and \, 
${\bf S} = (S_{1}\,,\,S_{2}\,,\,S_{3})$  
where  
 
\begin{equation} 
a_{S_{1}} = a_{+ -} + a_{- +} \;, 
\eqnum{3.7a} 
\end{equation} 
 
\begin{equation} 
a_{S_{2}} = i(a_{+ -} - a_{- +})\;, 
\eqnum{3.7b} 
\end{equation} 
 
\begin{equation} 
a_{S_{3}} = a_{+ +} - a_{- -} \;, 
\eqnum{3.7c} 
\end{equation} 
 
\noindent 
the equation (3.1) can be written in a more simple way and it reads 
 
\begin{equation} 
({\bf \nabla} \times {\bf a}) .\, {\bf S} = 1 \; . 
\eqnum{3.8} 
\end{equation} 
 
The form of the differential equation (3.8) is equal to that obtained  
in Refs.[9,10]. 
Then, the fact that the kinetic term can be written  
as a function of a vector field ${\bf {a}}$ which satisfies  
the equation (3.8) is recovered. It must be noted that the equation (3.8)  
is a good definition for a curl on a $S^2$ manifold. Then, the equation (3.8)  
together with (2.15) written in terms of the new variables $S_{1}\,,\, 
S_{2}$ and $S_{3}$, 
allows us to write the kinetic term in the Lagrangian as the area of a  
sphere with $\beta^{1/2}$ radious.       
This is the principal argument to say that $\beta^{1/2}$ must be integer 
or half-integer. For a complete discussion about this argument see Refs.[9,10]. 
 
\section{A simple case and its relation with previous works} 
 
From  section III, we can assert that a big family 
of Lagrangian really exists and any one of them can be considered as a good  
candidate for describing the dynamics contained in the commutation rules 
of the X-operators. The aim of this section is to discuss some important 
points by using explicitly one of the possible Lagrangians found in the  
previous section. 
Thus, by taking $a_{++} = a_{--} = 0$ in equation (3.4c) and calling 
$\alpha = -2s$ and $\beta = s^{2}$, the Lagrangian (2.10 can be written 
 
\begin{equation} 
L(X,\dot{X}) = \frac{i}{2} \left(\frac{X^{- +}\dot{X^{+ -}} -        
X^{+ -}\dot{X^{- +}}}{s + \frac{1}{2}(X^{+ +} - X^{- -})}\right) - H(X)\;, 
\eqnum{4.1} 
\end{equation} 
 
\noindent 
with the two constraints 
 
\begin{equation} 
X^{+ -} X^{- +} + \frac{1}{4}(X^{+ +} - X^{- -})^{2} = s^{2}\;, 
\eqnum{4.2a} 
\end{equation} 
 
\begin{equation} 
X^{+ +} + X^{- -} = 1\;. 
\eqnum{4.2b} 
\end{equation}

The equations (4.1) and (4.2) describe the classical dynamics of a system 
in which the commutation rules (1.1) are verified. 
 
We note that the same result also can be found by using the Dirac theory  
for constrained systems\cite{7}. 
From this approach it is easy to show that the constraints given in equations 
(4.2) together with the constraints coming from the definition  
of the momentum of the $X$'s variables, is  
a set of second class constraints. The Dirac brackets associated to  
this set of constraints are exactly the correct commutation rules for the  
Hubbard operators.     
 
Now, we are able to write the partition function by using the 
path integral Faddeev-Senjanovic approach\cite{11} and it reads  
 
\begin{eqnarray} 
Z & = & \int DX\; \delta[X^{+-}X^{-+}+\frac{1}{4}(X^{++}-X^{--})^2-s^2]\; 
\delta(X^{+ +} + X^{- -} - 1) \nonumber \\  
& \times & exp\;i \int dt\; L(X,\dot{X}) \;, 
\eqnum{4.3} 
\end{eqnarray} 
 
\noindent 
where $L(X,\dot{X})$ is given by (4.1). 
 
By integrating in the $X^{- -}$ variable we obtain for the partition 
function $Z$ the following expression 
 
\begin{eqnarray} 
Z & = &\int DX^{- +} DX^{+ -} DX^{+ +}\; \delta[X^{+ -}X^{- +}+  
\frac{1}{4}(2X^{+ +} - 1)^2] \nonumber \\ 
 & \times & exp\;i \int dt\; L^{(1)}(X,\dot{X})\;, 
\eqnum{4.4} 
\end{eqnarray} 
 
\noindent 
where  
 
\begin{equation} 
L^{(1)}(X,\dot{X})= - \frac{i}{2} \frac{X^{+-} \dot{X^{-+}} - X^{-+} 
\dot{X^{+-}}} 
{s + \frac{1}{2}(2X^{++}-1)} 
-H(X)\;. 
\eqnum{4.5} 
\end{equation} 
 
Making in equation (4.5) the following change of variables,

\begin{equation} 
S_{1} = \frac{X^{+ -} + X^{- +}}{2}\;, 
\eqnum{4.6a} 
\end{equation}

\begin{equation} 
S_{2} = \frac{X^{+ -} - X^{- +}}{2i}\;, 
\eqnum{4.6b} 
\end{equation}

\begin{equation} 
S_{3} = \frac{1}{2}(2 X^{++} - 1)\;, 
\eqnum{4.6c} 
\end{equation}

\noindent 
the functional integral (4.4) can be written as 
 
\begin{equation} 
Z= \int DS \;\delta(S_{1}^2 + S_{2}^2 + S_{3}^2 - s^2)\; exp\;i \int dt  
\;L^{(2)}(S,\dot{S})\;, 
\eqnum{4.7} 
\end{equation} 
 
\noindent  
where 
 
\begin{equation} 
L^{(2)}(S,\dot{S})=\frac{S_2 \dot{S}_1 - S_1 \dot{S}_2}{s+S_3} - H(S) 
\eqnum{4.8} 
\end{equation} 
 
\noindent 
where the constant Jacobian of the transformation (4.6) was absorbed in the  
functional integral measure. Therefore, the equation (4.7) for the partition 
function agrees with the 
expression (3.17) of Ref.[12], obtained by means of different arguments. 
 
Now, it is easy to show that this expression is consistent with the  
quantization of a spin system in the limit of large $s$. 
Applying again the Dirac theory, but now to the Lagrangian  
(4.8) with the constraints  
 
\begin{equation} 
{\mid{\bf{S}}\mid}^2 = s^2 
\eqnum{4.9}\;, 
\end{equation} 
 
\noindent 
we find that the second class nature of the constraint defining  
Dirac brackets are again exactly the commutation rules (1.1) for the  
spin components. 
It is interesting to note that in the quantization procedure, the second  
class constraint (4.9) must be considered as a strong equation  
among operators. Then, 
 
\begin{equation} 
{\hat{S}}^2 = s^2{\hat{I}} 
\eqnum{4.10}\;. 
\end{equation} 
 
From the comment given at the end of section III, it is known that the 
number $s$ 
must be integer or half-integer. Consequently, in the equation (4.10) it is  
not possible to write  $s^2$ as $s^{'}(s^{'}+1)$ with $s^{'}$ integer or  
half-integer. 
   
This is one of the important reasons which allows to ensure that  
in the path integral formalism for the spin systems, the information  
of large $s$ approximation is contained from the begining. This fact is  
connected with our 
results making impossible the inclusion of the full $X$-operators algebra  
in a classical Lagrangian formalism, or equivalently in a path integral  
formulation.

\section{Diagrammatic and Feynman rules} 
 
Now, in order to obtain the  
diagrammatic and the Feynman rules for the model the perturbative 
treatment is analyzed.  
The starting point is to consider the following partition function  
 
\begin{eqnarray} 
Z & = & \int DX^{+-}\;DX^{-+}\;Du \;\delta(X^{+-}X^{-+}+\frac{1}{4}u^{2} - 
\beta)\; \nonumber \\  
& \times & exp\;i \int dt\; L(X,\dot{X}) \;, 
\eqnum{5.1} 
\end{eqnarray} 
 
\noindent 
where the integration on the $(X^{++} + X^{--})$ variable has been made  
by using the function $\delta(X^{+ +} + X^{- -} - 1)$. 
 
Consequently, the Lagrangian $L(X,\dot{X})$ can be written: 
 
\begin{equation} 
L = a_{+-}{\dot{X}}^{+-} + a_{-+}{\dot{X}}^{-+} + a_{u}{\dot{u}} -  
H(X^{+-},X^{-+},u) \;. 
\eqnum{5.2} 
\end{equation}  
 
Taking into account the equations (3.4) for the  
coefficients, we consider the perturbative development for large value of  
the parameter $\beta$. Therefore the non-polynomial Lagrangian (5.2), up to  
first order in $\beta^{-1}$ reads: 
 
\begin{equation} 
L(X,\dot{X})=  \frac{i\alpha}{4\beta} (X^{-+} \dot{X^{+-}} - X^{+-} 
\dot{X^{-+}}) + a_{u}\dot{u} + \frac{i}{4 \beta} u (X^{-+} \dot{X^{+-}} -  
X^{+-}\dot{X^{-+}}) - H(X)\;. 
\eqnum{5.3} 
\end{equation} 
 
In equation (5.3), we consider for the Hamiltonian $H(X)$ the Heisenberg 
ferromagnetic form:  
 
\begin{equation} 
H(X) = - \frac{1}{2} J (X^{+-} X^{-+} + X^{-+} X^{+-} + \frac{1}{2}u u)\;. 
\eqnum{5.4} 
\end{equation} 
 
\noindent 
where $J>0$. 
 
By using in the path integral (5.1) the Gaussian representation  
 
\begin{eqnarray*} 
\delta(x) = \lim_{\sigma \rightarrow 0}{\frac{1}{\pi \sqrt{\sigma}}}\;  
exp(- \frac{1}{\sigma}\;x^{2})\;  
\end{eqnarray*} 
 
\noindent 
for the delta function, the partition function can  
be written in terms of an effective Lagrangian as follows 
 
\begin{equation} 
Z  =  \int DV\; exp\;i\; \int_{0}^{T}\; dt\; L^{eff}(V) \;. 
\eqnum{5.5} 
\end{equation} 
  
In the equation (5.5), the effective Lagrangian $L^{eff}(V)$ is written in 
terms of an extended complex vector field $V$ whose components are given by  
 
\begin{eqnarray*} 
V = (X^{+-}\;,\;X^{-+}\;,\;u)\;,  
\end{eqnarray*} 
 
\noindent 
and it can be partitioned as follows    
 
\begin{equation} 
L^{eff}  = L^{(2)}(V) + L^{(3)}(V) + L^{(4)}(V)\; . 
\eqnum{5.6} 
\end{equation} 
 
As it is usual the quadratic part $L^{(2)}(V)$ of the effective Lagrangian 
defines the free propagator of the model, and the remaining parts 
$L^{(3)}(V)$ and $L^{(4)}(V)$ represent the interaction vertices, i.e the  
three and 
four legs vertices of the model respectively. So, from equation (5.5) it 
can be seen that the quantum problem remains defined in terms of a path 
integral which contains the three independent fields 
$X^{+-}\;,\;X^{-+}$ and $u$. 
 
In equation (5.6) the quadratic part $L^{(2)}(V)$ is given by,

\begin{equation}  
L^{(2)}(V) = \frac{1}{2} V^{\alpha} D_{\alpha \beta} V^{\beta}\;, 
\eqnum{5.7} 
\end{equation} 
 
\noindent 
where  
 
\begin{eqnarray} 
D_{\alpha \beta} = 
\left( \matrix { 
0  &\frac{i \alpha}{4 \beta}\partial_{t} + \frac{\beta}{\sigma} + J  &0   \cr 
- \frac{i \alpha}{4 \beta}\partial_{t} + \frac{\beta}{\sigma} + J  &0  &0  \cr 
0     &0   &a \partial_{t} + \frac{\beta}{2 \sigma} + \frac{J}{2}   \cr } 
\right) \; . \eqnum{5.8} 
\end{eqnarray} 
 
The simplest case in which $a_{u} = h(u) = a u$  
(where $a$ is an arbitrary constant) has been considered when the  
matrix (5.8) was computed.   
 
The matrix $D_{\alpha \beta}$ is Hermitian and non-degenerate, and so the  
propagador $(D_{\alpha \beta})^{-1}$ in the $[q,\omega]$ - space can be 
evaluated and it results,  
 
\begin{eqnarray} 
(D_{\alpha \beta})^{-1}(\omega,\omega^{'}) = 
\left( \matrix { 
0  &\frac{4 \beta}{\alpha(\omega + \frac{4 \beta^{2}}{\alpha \sigma} -  
\frac{4 \beta}{\alpha} J_{q})}        &0                      \cr 
\frac{4 \beta}{\alpha(- \omega + \frac{4 \beta^{2}}{\alpha \sigma} -  
\frac{4 \beta}{\alpha} J_{q})}            &0                &0  \cr 
0     &0   &\frac{1}{i a \omega + \frac{\beta}{2 \sigma} - J_{q}}   \cr } 
\right)\delta(\omega,\omega^{'})  \; . \eqnum{5.10} 
\end{eqnarray} 
 
We note that $J_{q}$ is the Fourier transforms of $J_{ij} = J$ only if $i,j$ 
are nearest neighbor sites. 
 
The three and four legs vertices are respectively given by the parts 
  
\begin{equation}  
L^{(3)}(V) = \frac{1}{3!}\lambda_{\alpha\beta\gamma} V^{\alpha} V^{\beta} 
V^{\gamma} \;,  
\eqnum{5.11} 
\end{equation} 
 
\begin{equation}  
L^{(4)}(V) = \frac{1}{4!}\lambda_{\alpha\beta\gamma\delta} V^{\alpha} V^{\beta} 
V^{\gamma}  V^{\delta} \;,  
\eqnum{5.12} 
\end{equation} 
 
\noindent 
where  
  
\begin{equation}  
\lambda_{\alpha\beta\gamma} = \frac{1}{4 \beta}(\omega^{'} - \omega) 
\delta(-\omega + \omega^{'} + \omega^{''}) \delta(-q + q^{'} + q^{''}) 
\hspace{1cm} \alpha\neq\beta\neq\gamma\;,  
\eqnum{5.13} 
\end{equation}

\begin{equation}  
\lambda_{\alpha\beta\gamma\delta} = - \frac{3}{2 \sigma} 
\delta(\omega + \omega^{'} + \omega^{''} + \omega^{'''})  
\delta(q + q^{'} + q^{''} + q^{'''}) \hspace{1cm}  
\alpha = \beta = \gamma = \delta = 3 \;, 
\eqnum{5.14} 
\end{equation}

\begin{eqnarray}  
\lambda_{\alpha\beta\gamma\delta} = - \frac{1}{\sigma} 
\delta(-\omega + \omega^{'} + \omega^{''} + \omega^{'''})  
\delta(-q + q^{'} + &q^{''}& + q^{'''})  
\hspace{1cm} \alpha = 1, \beta = 2, \gamma = \delta = 3 \nonumber \\ \; 
             &and&\; all\; the\; permutations, 
\eqnum{5.15} 
\end{eqnarray} 
 
\noindent

\begin{eqnarray}  
\lambda_{\alpha\beta\gamma\delta} =  - \frac{4}{\sigma} 
\delta(-\omega + \omega^{'} - \omega^{''} + \omega^{'''})  
\delta(-q + q^{'} - &q^{''}& + q^{'''}) 
\hspace{1cm}  \alpha = 1, \beta = 2, \gamma = 1 \nonumber \\ \; 
            &and& \; all\; the\; permutations. 
\eqnum{5.16} 
\end{eqnarray}

From the above results we can see that for $\alpha = - 2 \sqrt{\beta} =  
- 2s$ and by choosing for the parameter $\sigma$ the value  
$\sigma = \frac{\beta}{Jz}$, where $z$ is the number of nearest neighbor sites, 
the matrix element reads 
 
\begin{equation}  
(D_{12})^{-1} \equiv \langle T[X_{q}^{+-}(\omega) X_{q^{'}}^{-+}(\omega^{'})] 
\rangle = \frac{2s}{\omega - 2sz(J - J_{q})}\;\delta(\omega - \omega^{'}) 
\delta(q - q^{'})\;.  
\eqnum{5.17} 
\end{equation} 
 
\noindent 
The above equation gives precisely the magnon propagator of the usual  
spin-wave theory.  
 
From (5.10), it can be seen that the longitudinal mode  
${\langle{T[u\;u]}\rangle}$ 
has a pole on the imaginary axis. This non-physical mode is related with 
the fact that there is no longitudinal dynamics in the lowest order of the 
spin-wave theory of the Heisenberg ferromagnetism. So, not losing physical  
information, also can be taken $a_{u} = 0$. 
 
By computing the propagator and vertices for the solution (3.5) at the same  
perturbative order, it is easy to show that the same results are obtained. 
In particular the free propagator takes the form (5.8) with $a =0$.   
 
In a further work under preparation, we will apply our 
perturbative approach to the renormalization and dumping of magnon energies.

\section{Conclusions and Discussions} 
 
In this paper a new discussion about the path integral formalism for 
dynamical systems written in terms of Hubbard operators is done. 
 
If it could have been possible, this path integral must have contained 
the full algebra (1.1)-(1.3) for the $X$-operators. 
Using the Faddeev-Jackiw symplectic formalism we have shown that this 
proposal is not possible, and in order to satisfy the commutation 
relations  rules (1.1) we have resigned to include the complete information 
contained in the $X$-operator algebra (1.1)-(1.3). 
 
By consistency of the formalism and in order to satisfy the  Hubbard  
commutation rules we have found the number of constraint conditions.  
From our point of view and in a total independent way we arrive to a  
path integral which is consistent with those obtained by means of      
coherent states method.  
 
We have also shown that this path integral for the spin system case,  
is valid in the large spin $s$ limit. Then, our conclussion assert that this  
limit is closely related with the impossibility to include the full algebra  
of the Hubbard $X$-operators in a classical dynamics. 
 
On the basis of our path integral formulation we present the 
diagrammatic and the Feynman rules for the perturbation theory. We have shown 
that our free theory is consistent with the results provided by the lowest  
order of the spin-wave theory. 
 
Finally, we can emphasize that from our approach a large family of kinetic  
terms of effective Lagrangians can be found. Some of them can be related with  
previous Lagrangians obtained by different methods.  
  
\vspace{0.5cm} 
 
Acknowledgements 
 
We acknowledge A. Dobry for many useful discussions. 
AG acknowledges parcial support on this project by Fundaci\'on 
Antorchas.

\end{document}